\documentclass[conference]{IEEEtran}
\IEEEoverridecommandlockouts
\usepackage{cite}
\usepackage{amsmath,amssymb,amsfonts}
\usepackage{algorithmic}
\usepackage{graphicx}
\usepackage{textcomp}
\usepackage{xcolor}
\usepackage[ruled,vlined]{algorithm2e}%
\usepackage[flushleft]{threeparttable}%
\usepackage{xcolor}
\usepackage{subcaption} 
\def\BibTeX{{\rm B\kern-.05em{\sc i\kern-.025em b}\kern-.08em
    T\kern-.1667em\lower.7ex\hbox{E}\kern-.125emX}}
\begin{document}

\title{Advanced Quantum Annealing Approach to \\ Vehicle Routing Problems with Time Windows\\
\thanks{James B. Holliday acknowledges support from J.B. Hunt Transport Inc.
Eneko Osaba acknowledges support from the Basque Government through the “Plan complementario de comunicación cúantica” (EXP.2022/01341) (A/20220551). }}

\author{\IEEEauthorblockN{James B. Holliday\IEEEauthorrefmark{1}}
\IEEEauthorblockA{\textit{CVIU Lab} \\
\textit{University of Arkansas}\\
\textit{J.B. Hunt Inc.}\\
Fayetteville, AR, USA \\
jbhollid@uark.edu}
\and
\IEEEauthorblockN{Darren Blount\IEEEauthorrefmark{1}}
\IEEEauthorblockA{\textit{CVIU Lab} \\
\textit{University of Arkansas}\\
Fayetteville, AR, USA \\
darrenb@uark.edu}
\and
\IEEEauthorblockN{Eneko Osaba}
\IEEEauthorblockA{\textit{TECNALIA, Basque Research and} \\ 
\textit{Technology Alliance (BRTA)}\\
48160 Derio Spain \\
eneko.osaba@tecnalia.com}
\and
\IEEEauthorblockN{Khoa Luu\IEEEauthorrefmark{1}}
\IEEEauthorblockA{\textit{CVIU Lab} \\
\textit{University of Arkansas}\\
Fayetteville, AR, USA \\
khoaluu@uark.edu}
\IEEEauthorblockA{
    \IEEEauthorrefmark{1}\tt\scriptsize{https://uark-cviu.github.io/}
    }
}

\maketitle

\begin{abstract}
In this paper, we explore the potential for quantum annealing to solve realistic routing problems. We focus on two NP-Hard problems, including the Traveling Salesman Problem with Time Windows and the Capacitated Vehicle Routing Problem with Time Windows. We utilize D-Wave's Quantum Annealer and Constrained Quadratic Model (CQM) solver within a hybrid framework to solve these problems. We demonstrate that while the CQM solver effectively minimizes route costs, it struggles to maintain time window feasibility as the problem size increases. To address this limitation, we implement a heuristic method that fixes infeasible solutions through a series of swapping operations. Testing on benchmark instances shows our method achieves promising results with an average optimality gap of 3.86\%. 
\end{abstract}


\section{Introduction} \label{intro}
To date, quantum computing (QC) has shown considerable promise for solving combinatorial optimization problems \cite{abbas2023quantum}. However, in the Noisy Intermediate-Scale Quantum (NISQ) era, QC has only shown this capability on problems with limited training data. Hybrid approaches to QC have shown that more significant realistic problems can be solved. Hybrid means that part of the problem is solved using classical computing, and part is solved using QC \cite{holliday2024tabu, XBacWK2, XBacWK1, XBacjournal1, AdityaQIP}. Separating the problem into classical and quantum components allows for scaling up the complexity of the problem from toy-sized to realistic.

\begin{figure}[t]
    \centering
    \includegraphics[width=0.98\linewidth]{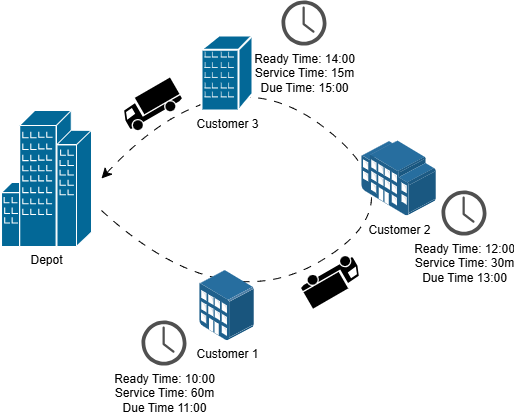}
    \caption{Vehicle Route with Time Window Constraints}
    \label{fig:timewindowroute}
\end{figure}

\textbf{The Contributions in this Work:} 
This work advances the Hybrid Quantum Tabu Search (HQTS) method, originally designed as a hybrid meta-heuristic for the Capacitated Vehicle Routing Problem (CVRP)~\cite{holliday2024tabu}. It will address the more complex Capacitated Vehicle Routing Problem with Time Windows (CVRPTW) as Fig. \ref{fig:timewindowroute}. Our primary contribution lies in adapting HQTS to incorporate time window constraints, a significant departure from the capacity-only focus of the Capacitated Vehicle Routing problem.

First, a key innovation is transitioning from a Quadratic Unconstrained Binary Optimization (QUBO) formulation, used in our CVRP work for distance minimization, to D-Wave's specialized Constrained Quadratic Model (CQM) solver. A CQM-based approach for the CVRPTW enables explicit modeling of time window constraints alongside travel cost objectives. While the QUBO formulation sufficed for capacity-constrained routing, the addition of time windows necessitated a shift to CQM to handle the interplay between distance minimization and temporal feasibility. 
 
Second, while the CQM solver efficiently minimizes distance, it sometimes fails to maintain time window feasibility. Recognizing the persistent feasibility challenges, we introduce a novel post-processing heuristic to repair time window violations, which enhances HQTS by correcting infeasible CQM outputs. This approach ensures practical applicability by guaranteeing feasible outcomes, a critical advancement over relying solely on the quantum solver’s unconstrained outputs.

Finally, our experimental evaluation using the Solomon Benchmark Dataset~\cite{solomon1987algorithms} further contributes by providing insights into the scalability and limitations of QA for time-constrained routing. For the TSPTW, we demonstrate that feasibility diminishes as the number of stops increases beyond thirteen, despite cost reductions, highlighting the CQM solver’s sensitivity to problem size and initial conditions. For the CVRPTW, updated HQTS achieves an average optimality gap of 3.85\% across selected tight-window instances, showcasing its effectiveness. Collectively, these contributions advance the application of quantum annealing to real-world logistics, offering a robust hybrid framework that balances optimization and feasibility under complex constraints.

This article is organized as follows: section \ref{background} reviews the background, section \ref{formulations} defines problems, section \ref{method} details our algorithm, section \ref{experiments} presents experimental setup and results, and section \ref{conclusion} finally offers conclusions.

\section{Background} \label{background}

This section briefly addresses two pivotal concepts to contextualize the current research. Firstly, in Section \ref{sec:problems}, we will concisely describe the problems addressed in this research. Secondly, in Section \ref{sec:QA}, we succinctly describe the quantum annealing paradigm and the hybrid solver considered in this work, i.e., D-Wave's CQM solver. We finish this section with a brief outline of the related studies.

\subsection{Routing Problems with Time Windows}\label{sec:problems}
Route planning and logistics problems hold significant importance in the scientific world due to their wide-ranging applications in various industries, including transportation, delivery services, and supply chain management. These problems can be mathematically modeled as the Traveling Salesman Problem (TSP) \cite{hoffman2013traveling} and the Vehicle Routing Problem (VRP) \cite{toth2002vehicle}. The TSP focuses on finding the shortest route that visits a set of cities and returns to the origin city. At the same time, the VRP extends this concept by considering multiple vehicles and routes to service a set of customers.

Although the TSP and VRP are interesting problems from an academic perspective, they often fall short when addressing real-world problems. Real-life scenarios typically involve additional constraints and complexities that the basic models do not capture. To bridge this gap, various constraints are introduced to create more realistic variants of these problems \cite{ilavarasi2014variants,caceres2014rich}. One of the most widely used constraints is to include time windows, adding a layer of practicality to the models \cite{kallehauge2005vehicle}.

Time windows are the constraints that specify the time intervals within which specific tasks must be completed. For example, in the context of delivery services, each customer may have a preferred time window when they are available to receive their delivery. Incorporating time windows into the TSP and VRP ensures the solutions are optimal in terms of distance or cost and feasible in meeting customer requirements. It makes the models more applicable to real-world scenarios, where timing and scheduling are crucial factors.

\subsection{Quantum Annealing and Constrained Quadratic Model (CQM)} \label{sec:QA}

A Quantum Annealer (QA) \cite{morita2008mathematical} is a specialized quantum device designed to tackle optimization problems through a process inspired by classical Simulated Annealing \cite{van1987simulated}. These devices utilize quantum mechanics to efficiently explore solution spaces, aiming to find the optimal value of a fixed objective function. Recent advancements in quantum technologies have led to the development of intermediate-scale QAs, which include quantum annealing for programmable applications. Among these, \textit{D-Wave Systems}' QAs, based on superconducting qubits, are the most widely used in areas such as finance \cite{bouland2020prospects}, logistics \cite{osaba2022systematic}, and industries \cite{yarkoni2022quantum}.

In this paper, the Quantum Annealing paradigm is presented to address the routing problem introduced earlier. More specifically, we use the hybrid method known as the \textit{LeapCQMHybrid} solver (LeapCQM), which is part of the D-Wave Hybrid Solver Service (HSS) \cite{HSS}. In a nutshell, the HSS portfolio encompasses a suite of hybrid heuristics that seamlessly integrate quantum and classical computation to address large-scale optimization problems and real-world industrial applications. As of this writing, the HSS includes four approaches, including the binary quadratic model (BQM) algorithm for problems formulated with binary variables; the discrete quadratic model (DQM) approach to problems described with integer variables; the constrained quadratic model (CQM) technique, the one used in this research and can handle problems defined with binary, integer, and even real variables; and the Nonlinear-Program Hybrid Solver, which excels at natively accommodating nonlinear (linear, quadratic, and higher-order) inequality and equality constraints, even when expressed arithmetically.

The workflow of the LeapCQM solver \cite{leapCQM} is divided into two phases. Initially, the input problem is read, and a set of parallel hybrid threads is built. Each thread comprises a Classical Heuristic Module (CM), responsible for exploring the entire solution space, and a Quantum Module (QM). The QM generates quantum queries that are transmitted to a back-end Advantage QPU. The responses received from the QPU help direct the heuristic module to more promising regions of the search space and can also enhance existing solutions. Subsequently, all branches are executed independently, and the best solution found from the entire set of threads is given to the user.

Finally, it is noteworthy that the QM module of the LeapCQM employs the \texttt{Advantage\_system7.1} device, which represents the latest advancement from D-Wave at the time of this writing. This quantum computer is equipped with 5,554 working qubits and over 35,000 couplers, organized in a Pegasus topology~\cite{boothby2020next}. 

\begin{figure*}[hbtp]
  \centering
  \fbox{\includegraphics[width=0.98\textwidth]{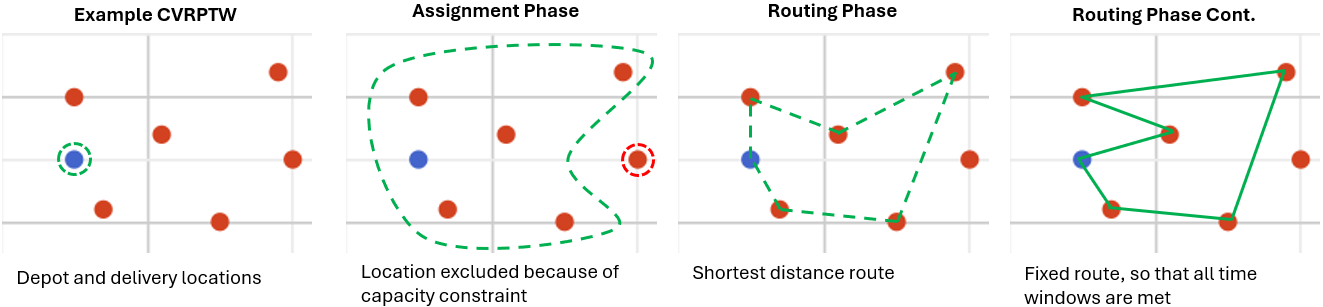}}
  \caption{Solving the CVRPTW with a phased approach}
  \label{fig:teaser}
\end{figure*}

\subsection{Related Work}

Recent studies have demonstrated the applicability of quantum annealing to various logistics and routing problems. For instance, hybrid quantum-classical algorithms have been developed to tackle the Capacitated Vehicle Routing Problem (CVRP), which involves optimizing routes while considering vehicle capacity constraints. These hybrid approaches combine the strengths of quantum annealing with classical heuristics to handle more significant problem instances and improve solution accuracy. Some examples of this specific trend can be found in \cite{sinno2023performance,irie2019quantum,feld2019hybrid}. In addition, research has shown that quantum annealing can be effectively used for real-time route optimization, addressing dynamic changes in logistics networks \cite{harikrishnakumar2020quantum}, and more advanced constraints such as priority clients or heterogeneous fleet of vehicles \cite{osaba2024solving}.

Regarding the TSP, this problem has been extensively studied from a quantum perspective. Due to its academic nature, it has been frequently used as a benchmarking problem \cite{qian2023comparative,osaba2025d}, that is, to test the effectiveness of novel solution methods. Nevertheless, although to a lesser extent, it has also been used to address real-world logistic problems \cite{weinberg2023supply,osaba2025solving}.

In particular, and pertinent to our study, time windows are the most extensively examined constraint from a classical standpoint. The critical role of time windows in logistics is clear, as fulfilling orders for various businesses or individuals require adherence to their specific schedules and availabilities. Although time windows have been theoretically or preliminarily explored from a quantum perspective in studies like \cite{irie2019quantum} and \cite{harwood2021formulating}, their application to medium or large-scale problems remains largely uncharted, with limited examples such as the work published in \cite{weinberg2023supply}. In that study, the authors address multi-truck vehicle routing for supply chain logistics. Since solving the problem using a fully embedded approach was not feasible, they proposed an algorithm that iteratively assigns routes to trucks. This approach allowed for the consideration of constraints such as restricted driving windows. Essentially, the lack of practical studies addressing time windows significantly enhances the originality of our research.

\section{Problem Formulation} \label{formulations}

\subsection{Traveling Salesman Problem with Time Windows (TSPTW)}

The TSPTW is a variation of the standard TSP, incorporating additional constraints that require each location or node to be visited within a specific time window. We define the binary variable \( x_{i,j} \) with the value of one if the vehicle travels directly from node \( i \) to node \( j \), and zero otherwise. We define the continuous variable \( z_{i,j} \) as the relative ordering of visits between nodes.

The primary goal or objective function of the optimization is to minimize the total travel cost of the route by putting the nodes into a sequence that leads to the lowest cost.
\begin{align}\label{objective_function}
\min \sum_{i \in \mathcal{N}} \sum_{j \in \mathcal{N}, i \neq j} \text{c}(i,j) \cdot x_{i,j}
\end{align}
\begin{align}\label{one_visit}
\sum_{i \in \mathcal{N}, i \neq j} x_{i,j} = 1, \quad \forall j \in \mathcal{N}
\end{align}
\begin{align}\label{one_route}
\sum_{j \in \mathcal{N}, j \neq i} x_{i,j} = 1, \quad \forall i \in \mathcal{N}
\end{align}
\begin{align}\label{subtour_one}
\sum_{j \in \mathcal{N}, j \neq i} z_{i,j} - \sum_{j \in \mathcal{N}, j \neq 0, j \neq i} z_{j,i} = 1, \quad \forall i \in \mathcal{N}, i \neq 0
\end{align}
\begin{align}\label{subtour_two}
z_{i,j} \leq (|\mathcal{N}| - 1) x_{i,j}, \quad \forall i,j \in \mathcal{N}, i \neq j, i \neq 0
\end{align}
Eqn. \eqref{objective_function} denotes the objective function to minimize the route cost. \( \mathcal{N} \) represents the set of nodes, i.e., customers and depot. \( x_{i,j} \) is the binary decision variable indicating if the vehicle travels from node \( i \) to node \( j \). Here \( \text{c}(i,j) \) denotes the cost function, the Euclidean distance between nodes \( i \) and \( j \).  The constraints on the TSPTW are all the remaining equations.  Eqn. \eqref{one_visit} is the constraint that each node must be visited exactly once. Eqn. \eqref{one_route} forces the model so that each node has exactly one node after in the route. Eqn. \eqref{subtour_one} and Eqn. \eqref{subtour_two} are the sub-tour elimination constraints or flow constraints.  

These four constraints collectively enforce that the depot, i.e., node 0, is both the starting and ending point of the route. Because the depot has exactly one incoming and one outgoing edge, the only feasible way to satisfy these constraints is for the route to start at the depot and return to it, making any other starting or ending point impossible. Thus, the constraints inherently guarantee the depot is the first and last stop in the route without additional explicit constraints.
\begin{equation}\label{time_window}
\begin{gathered}
\text{if} \quad \text{ready\_time}_i + \text{service}_i + \text{cost}(i,j) \geq \text{due\_time}_j, \quad\\ \text{then} \quad z_{i,j} - z_{j,i} \geq 0\\
\end{gathered}
\end{equation}
To enforce time feasibility, we introduce the constraint in Eqn. \ref{time_window}. Here \( \text{ready\_time}_i \) is the earliest time the vehicle can depart from node \( i \). The vehicle \( \text{due\_time}_j \) is the latest time the vehicle must arrive at node \( j \). The \( \text{service}_i \) is the required service time at node \( i \). The \( \text{cost}(i,j) \) represents the travel time from node \( i \) to node \( j \). This constraint ensures that a vehicle does not travel to node \( j \) if the earliest departure time from \( i \), plus the travel and service time, would exceed \( j \)'s due time. If this condition is violated, the constraint enforces a reordering of visits to maintain feasibility.

\subsection{Capacitated Vehicle Routing Problem with Time Windows (CVRPTW)}

The CVRPTW generalizes the TSPTW by adding multiple routes or vehicles and constraints. The additional constraint is the capacity constraint of the vehicle. The objective function remains the same as the time windows constraint from the TSPTW, except that they are now extended for each vehicle. 
\begin{align}\label{capacity_constraint}
\sum_{i\in V} q_i \sum_{j\in V} x_{ijk} \leq Q \quad \forall k \in K
\end{align}
The constraint in Eqn. \eqref{capacity_constraint} limits nodes that can be added to vehicle \(k\), so that the capacity is not exceeded. Similar to before \(x_{ijk}\) is the binary variable representing that on vehicle \(k\) node \(i\) is visited before \(j\). \(Q\) is the capacity of the vehicle and \(q_i\) is the demand for node \(i\). 

\begin{figure}[tb]
    \centering
    \includegraphics[width=0.98\linewidth]{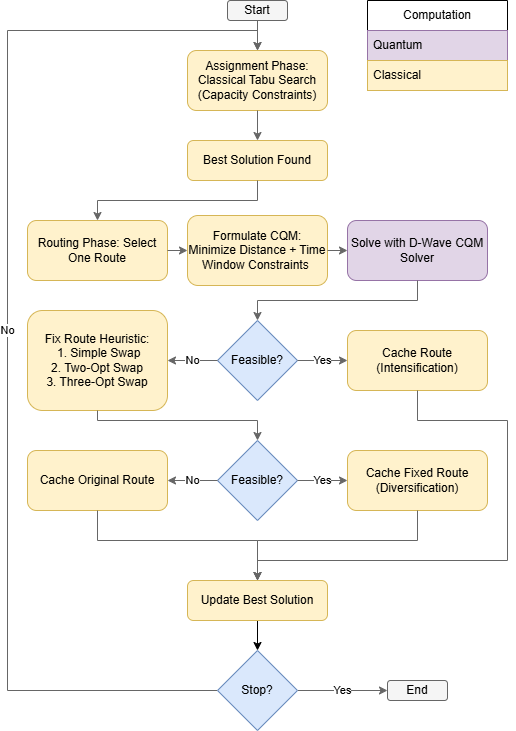}
    \caption{The proposed flow of HQTS for CVRPTW}
    \label{fig:tsflow}
\end{figure}

\section{Our Proposed Method} \label{method}
In this research, we advance the previous work by creating a hybrid meta-heuristic route optimizer named Hybrid Quantum Tabu Search (HQTS)\cite{holliday2024tabu}. For brevity, we leave the implementation of HQTS to our previous publication. In this work, we will focus on how we extended HQTS to be able to solve the CVRPTW. 

We can break the CVRPTW into two phases: an assignment phase, where we determine which locations should be on a given route, and a routing phase, where we decide the sequence of locations that should be visited along the route. Simply put, HQTS solves the assignment phase using a classical tabu search implementation and the routing phase utilizing the CQM solver. Refer to Fig. \ref{fig:teaser} for a visualization of the phased approach.

In our original approach to solving the CVRP, we only need to consider the capacity constraint when we assign the locations. It means a feasible route must ensure the sum of demand across all locations of the route stays at or under the capacity limit of the vehicle. When we go to the routing part, we only need to minimize the distance of the route; we already know the capacity is feasible. However, when we add time windows as another constraint, we must consider them when we do the routing because just minimizing the route's distance may, in fact, cause it to break a time window and create an infeasible route. 

Moving on from the assignment part, for solving the routing part of our algorithm we formulated a quadratic unconstrained binary optimization problem (QUBO). The goal of this part is to minimize the distance of the route by sequencing the locations on the route. It is essentially solving the TSP for one route. We provided the QUBO to D-Wave's QA to generate solutions. As we move to our new method for solving the CVRPTW we moved to the CQM solver instead of QUBO. However, in section \ref{experiments}, we discovered that the CQM solver was not universally successful at maintaining time window feasibility.   

The time window constraint creates a practical challenge for route optimization. Our mathematical formulation for solving the TSPTW within the larger CVRPTW underwent multiple refinements to balance time window feasibility with route cost minimization. Initially, we investigated a QUBO formulation, but transitioning to the CQM solver framework allowed for more explicit constraint and objective modeling. This shift improved interpretability; however, ensuring feasibility remained a significant challenge, particularly for larger problem instances.

To improve feasibility, arrival time variables were introduced so that sequencing constraints could be directly modeled rather than inferred through penalties. Arrival times were defined as real-valued variables rather than binary, allowing them to take continuous values within a feasible range. This provided a more precise representation of vehicle movements, enabling direct enforcement of constraints that ensured a vehicle could only depart a location after completing its service and arriving at the next node within the required time window. Despite this improvement, feasibility was not consistently achieved. The solver frequently produced infeasible solutions, particularly as the number of stops increased. 

Further refinement of the model involved scaling the cost matrix to normalize values and potentially improve solver performance. The rationale was that reducing the numerical disparity between travel costs might encourage the solver to search the solution space more effectively. However, this approach did not yield improved results. The solver struggled to maintain feasibility for more extensive routes, and scaling the costs had minimal impact on its ability to enforce time windows. 

An alternative formulation introduced sequencing-based variables, denoted as z, to define the relative ordering of visits explicitly. This approach proved more effective than direct penalization, allowing the solver to systematically account for precedence relationships between nodes. By comparing arrival times and enforcing an explicit order, the likelihood of generating infeasible solutions was reduced. The key improvement stemmed from restructuring time sequencing constraints so that the model enforced an ordering relationship based on their respective time windows for each pair of locations. The formulation attempted to ensure that if one node’s earliest possible departure time, combined with service and travel time, exceeded the due time of another, the solver would adjust the order of visits accordingly. This method produced the best results in terms of feasibility across multiple test cases, as it directly incorporated the precedence constraints rather than relying on penalties or implicit inference. However, feasibility was still not guaranteed in all cases, particularly as the number of nodes increased.

Ultimately, while the CQM framework allowed for a more structured approach to constraint modeling than QUBO, no single formulation consistently maintained feasibility across all instances. The introduction of z variables yielded the most promising results, yet heuristic post-processing was still necessary to correct infeasible solutions. Further refinement of the penalty-based approach might have improved feasibility, but the iterative nature of tuning made it impractical for real-world deployment. 

As we moved beyond solving the TSPTW and into the CVRPTW, we determined we needed to address the infeasibility limitation. Utilizing the QA was central to our research goal, so we decided to "fix" the results the CQM solver provided when it generated an infeasible route. We did this by evaluating the CQM solver results for time window feasibility before we cached them. If the route was infeasible, we ran simple swapping heuristics on the resulting route until it became feasible. We implemented three swapping methods: simple swap, two-opt swap, and three-opt swap \cite{croes1958method}. If the simple swapping method failed to create a feasible route, we would try the two-opt swap method. Again, if a two-opt swap failed to create a feasible route, we would try a three-opt swap method. If the route could be fixed at any point during these methods, we would save that route in the cache, and if the heuristics failed, we would save the original route before any optimization was attempted in the cache. Algorithm \ref{FRH} shows how the heuristic function works.

\begin{algorithm}
\DontPrintSemicolon
\KwData{$start=$ One route from the best solution found.}
\KwResult{$RouteCache=$ optimized route or fixed route.}
\Begin{
$r \longleftarrow CQMResults(start)$\;
\eIf{$IsFeasible(r)$}{$AddToRouteCache(r)$}
{\eIf{$SimpleSwapFixesRoute(r)$}{$AddToRouteCache(r)$}
{\eIf{$TwoOptSwapFixesRoute(r)$}{$AddToRouteCache(r)$}
{\eIf{$ThreeOptSwapFixesRoute(r)$}{$AddToRouteCache(r)$}
{$AddToRouteCache(start)$}}}}}
\caption{Fix Route Heuristic\label{FRH}}
\end{algorithm}

The heuristics created a novel intensification/diversification strategy for our method. If the CQM solver created a feasible and optimized route leading to a better solution we would have intensified the already best found solution. If the CQM solver failed to create a feasible route, we would take the optimal but infeasible sequence returned from the CQM solver and then modify it with the heuristics. It might not lead to an improved solution, but it would provide a feasible solution we might not have seen before, thus diversifying our search. Our results from this method are presented in the next section. Again, we refer to Fig. \ref{fig:tsflow} to visualize this strategy.

\section{Experiments and Results} \label{experiments}
\subsection{Datasets and Setup}
For our experiments, we utilized the Solomon Benchmark Dataset \cite{solomon1987algorithms} to model real-world delivery constraints. This dataset features 100 customer instances, each with a depot (node 0) where vehicles start and end, and customers are defined by X-Y coordinates, demand, ready time, due date, and service time. Time windows (ready time to due date) and vehicle capacity constraints ensure realistic scheduling and routing challenges, requiring vehicles to arrive within windows and manage limited capacity efficiently.

The dataset is split into six categories of problems. Clustered with tight time windows, clustered with open time windows, random with tight time windows, random with open time windows, and finally random and clustered with tight time windows, random and clustered with open time windows. A breakdown of the different types of problems is presented in Table \ref{dataset}.

\begin{table}[h]
\centering
\begin{threeparttable}    
    \caption{Solomon Dataset Problems}
    \label{dataset}
    \centering
        \begin{tabular}{llll}
        {Problem Range} & {No. of Locations} & {Orientation} & {Time Window}\\
        \hline
         C101 - 109 & 100 & C & T \\
         C201 - 208 & 100 & C & L \\
         R101 - 112 & 100 & R & T \\
         R201 - 211 & 100 & R & L \\
         RC101 - 108 & 100 & R and C & T \\         
         RC201 - 208 & 100 & R and C & L \\              
         \hline
    \end{tabular}

    \begin{tablenotes}
      \item C = Clustered, R = Random, T = Tight, L = Loose
    \end{tablenotes}
\end{threeparttable}   
\end{table}  

All TSPTW experiments were run on an ASUS ZenbookProDuo with an Intel Core i9-11900H processor and 32GB RAM in Visual Studio Code. For the CVRPTW, all experiments were run on a GitHub Codespace \cite{githubcodespaces}. The Codespace had a four-core processor with 16 GB of RAM that was used for the classical computation of our algorithm. The CQM solver for both experiments was D-Wave's LeapCQM, which has a QM that runs on the Advantage System 7.1 device. We accessed this system via D-Wave’s cloud API \cite{dwaveleap}.

\subsection{TSPTW}
The TSPTW was evaluated using the CQM solver to assess its ability to optimize travel costs while satisfying time constraints. These experiments examined how the number of stops, variations in initial cluster costs, and problem instance characteristics influenced optimization performance and solution feasibility.

\begin{figure}[b]
  \centering
  \includegraphics[width=\linewidth]{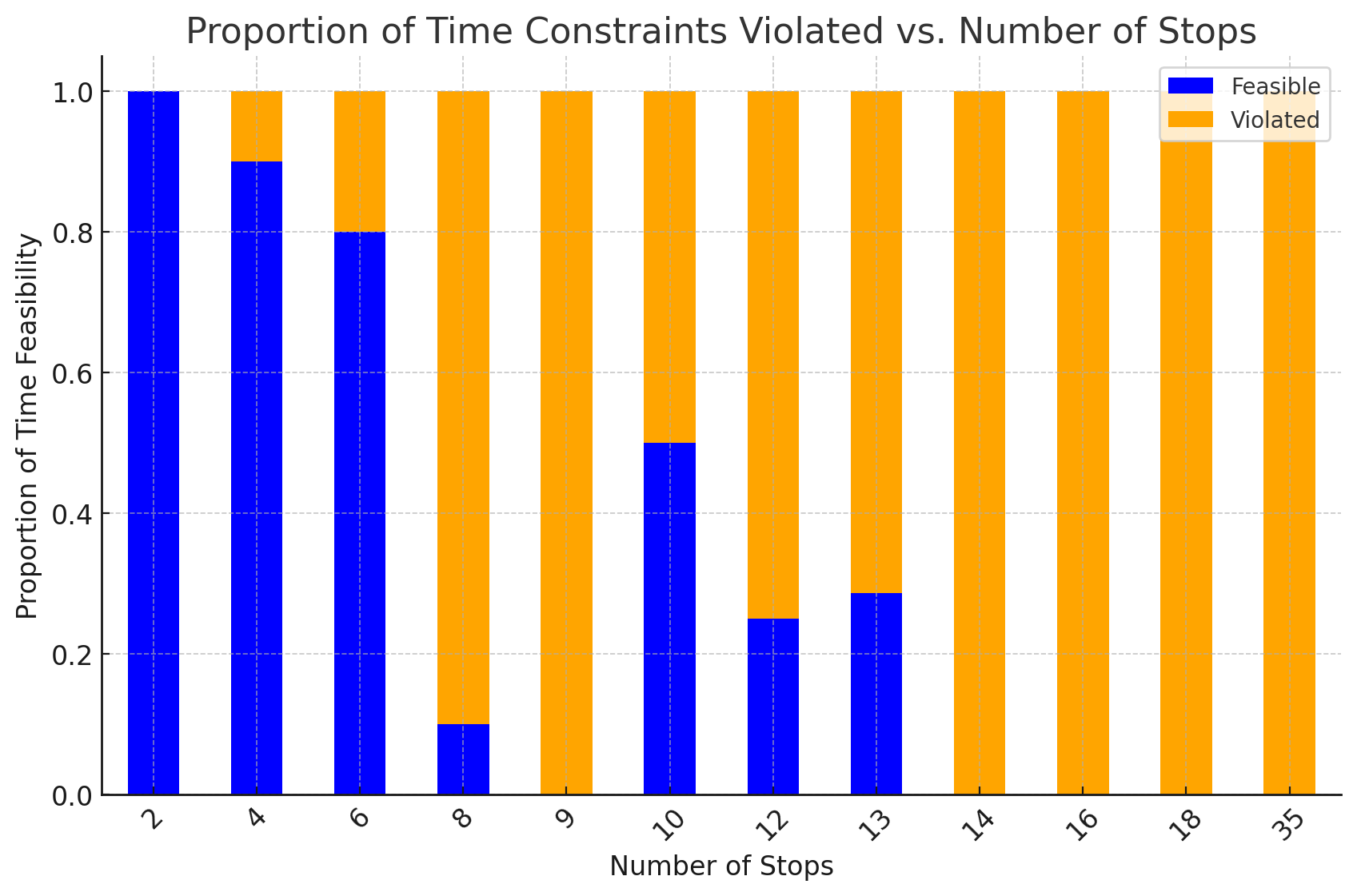}
  \caption{Proportion of Time Constraints Violated for Increasing Number of Stops}
  \label{fig:time_violations}
\end{figure}

Across tests from C207, C104, C102, and C202, results showed that as the number of stops increased, the number of feasible solutions decreased. The graphical analysis presented in Figure \ref{fig:time_violations} illustrates this trend, where small-scale problems generally maintained feasibility. Still, feasibility decreased significantly for larger-scale problems, particularly at thirteen and thirty-five stops.

For C207, initial route costs were varied by rearranging nodes within a route, leading to different optimization outcomes. In ten-stop routes, the initial costs ranged from 98.8 to 232.2, with optimized costs consistently reducing to approximately 98.8. Time feasibility was generally maintained, except in cases where the starting cost was significantly higher. Table \ref{tab:c207_10} summarizes the results for ten-stop routes.

\begin{table}[h]
\centering
\begin{threeparttable}    
    \caption{Optimization Results for C207 with 10 Stops}
    \label{tab:c207_10}
    \centering
        \begin{tabular}{lllllll}
        {Pre-Optimized} & {Optimize} & {Time} & {Improved} \\
        {Cost} & {Cost} & {Violated} & {Cost} \\
        \hline
        61.4 & 60.4 & Yes & Yes\\
        61.4 & 60.4 & Yes & Yes\\
        61.4 & 60.4 & Yes & Yes\\
        98.8 & 98.8 & No & No\\
        205.3 & 98.8 & No & Yes\\
        \hline
    \end{tabular}
\end{threeparttable}   
\end{table}

In thirteen-stop routes, the variation in initial costs was more pronounced, ranging from 100.1 to 276.6, resulting in inconsistent feasibility. Higher initial costs, such as 272.9, led to an optimized cost of 116 while maintaining feasibility, whereas a lower initial cost of 100.1 resulted in an optimized cost of 117.2 but violated time constraints. Some mid-range initial costs, such as 223.1 and 181.8, produced optimized solutions that remained within time constraints, indicating that initial conditions influenced feasibility. Table \ref{tab:c207_13} presents these results.

\begin{table}[h]
\centering
\begin{threeparttable}    
    \caption{Optimization Results for C207 with 13 Stops}
    \label{tab:c207_13}
    \centering
        \begin{tabular}{lllllll}
        {Pre-Optimized} & {Optimize} & {Time} & {Improved} \\
        {Cost} & {Cost} & {Violated} & {Cost} \\
        \hline
        100.1 & 117.2 & Yes & No\\
        223.1 & 110.7 & No & Yes\\
        272.9 & 116 & No & Yes\\
        181.8 & 101.5 & No & Yes\\
        246 & 120.6 & Yes & Yes\\
        \hline
    \end{tabular}
\end{threeparttable}   
\end{table}

In thirty-five-stop routes, feasibility was never achieved, regardless of the starting route cost. While significant cost reductions were observed, such as reducing the initial cost of 1134.4 to 938.6, none of the solutions satisfied the imposed time windows. These findings suggest that while the solver remains sensitive to initial conditions in moderate-sized problems, it cannot maintain feasibility as the problem scale increases. Table \ref{tab:c207_35} summarizes these results.

\begin{table}[h]
\centering
\begin{threeparttable}    
    \caption{Optimization Results for C207 with 35 Stops}
    \label{tab:c207_35}
    \centering
        \begin{tabular}{lllllll}
        {Pre-Optimized} & {Optimize} & {Time} & {Improved} \\
        {Cost} & {Cost} & {Violated} & {Cost} \\
        \hline
        235.4 & 841.6 & Yes & No\\
        1,049.6 & 1,048.1 & Yes & Yes\\
        1134.4 & 938.6 & Yes & Yes\\
        814.8 & 830.8 & Yes & No\\
        \hline
    \end{tabular}
\end{threeparttable}   
\end{table}

For C102, the solver produced different optimization outcomes despite maintaining the same number of stops, as initial route costs varied across runs. Unlike C207, where feasibility was preserved in some cases, C102 consistently resulted in time violations for all thirteen-stop routes. The initial route costs ranged from 67.6 to 101.2, but feasibility was not maintained in every instance. Table \ref{tab:c102_13} summarizes these findings.

\begin{table}[h]
\centering
\begin{threeparttable}    
    \caption{Optimization Results for C102 with 13 Stops}
    \label{tab:c102_13}
    \centering
        \begin{tabular}{lllllll}
        {Pre-Optimized} & {Optimize} & {Time} & {Improved} \\
        {Cost} & {Cost} & {Violated} & {Cost} \\
        \hline
        72.8 & 72.8 & Yes & No\\
        68.2 & 68.2 & Yes & No\\
        101.2 & 80.4 & Yes & Yes\\
        101.2 & 74.6 & Yes & Yes\\
        67.6 & 67.6 & Yes & No\\
        81.5 & 73.3 & Yes & Yes\\
        \hline
    \end{tabular}
\end{threeparttable}   
\end{table}

In the C102 dataset, the CQM solver demonstrated full determinism up to 10 stops. For each instance where the initial route was the same, the resulting route cost, optimized cost, time violations, and improvement status remained identical across all trials. This consistency indicates that the CQM solver operated deterministically within this range, yielding the same optimized cost and constraint violations for identical inputs. Testing was limited due to quantum computing resources to further these tests to routes with more stops. The results are summarized in Table \ref{tab:c102_consistency}.

\begin{table}[h]
\centering
\begin{threeparttable}    
    \caption{Consistency in C102 for up to 10 Routes}
    \label{tab:c102_consistency}
    \centering
    \begin{tabular}{lllllll}
         & {Pre-Optimized} & {Optimize} & {Time} & {Improved} \\
        {Stops} & {Cost} & {Cost} & {Violated} & {Cost} \\
        \hline
        2 & 38.9 & 38.9 & No & No \\
        2 & 38.9 & 38.9 & No & No \\
        2 & 38.9 & 38.9 & No & No \\
        4 & 44.2 & 44.2 & No & No \\
        4 & 44.2 & 44.2 & No & No \\
        4 & 44.2 & 44.2 & No & No \\
        6 & 50.0 & 50.0 & No & No \\
        6 & 50.0 & 50.0 & No & No \\
        6 & 50.0 & 50.0 & No & No \\
        8  & 56.7 & 54.0 & Yes & Yes \\
        8  & 56.7 & 54.0 & Yes & Yes \\
        8  & 56.7 & 54.0 & Yes & Yes \\
        9  & 70.8 & 59.3 & Yes & Yes \\
        9  & 85.5 & 59.3 & Yes & Yes \\
        10 & 61.4 & 60.4 & Yes & Yes \\
        10 & 61.4 & 60.4 & Yes & Yes \\
        10 & 61.4 & 60.4 & Yes & Yes \\
        \hline
    \end{tabular}
\end{threeparttable}   
\end{table}

Overall, the results demonstrate that while the CQM solver effectively reduces travel costs in small to medium-sized problems, feasibility becomes increasingly inconsistent as the number of stops grows. Optimization was generally successful for problems with ten or fewer stops, with deterministic behavior observed in certain instances. However, feasibility deteriorated significantly beyond thirteen stops, with larger problem instances (e.g., thirty-five stops) consistently failing to satisfy time constraints despite cost reductions. The solver’s sensitivity to initial conditions further influenced feasibility, suggesting that the starting route cost is critical in determining whether constraints can be met.

\subsection{CVRPTW}

For the next part of our experimentation, we deployed the CQM solver within HQTS, targeting six challenging instances from the Solomon dataset~\cite{solomon1987algorithms}. We deliberately excluded clustered problems (C1XX, C2XX), as the Clarke and Wright Savings algorithm~\cite{clarke1964scheduling}—used for HQTS’s initial solutions—often yields near-optimal results, leaving little room for quantum enhancement. Similarly, we bypassed loose time window problems (2XX), which typically feature few routes with many stops; our TSPTW findings indicated that CQM struggles with feasibility beyond 13 stops, rendering these cases less informative. Instead, we focused on three random tight problems (R101–R103) and three random-clustered tight problems (RC101–RC103), balancing complexity and relevance. Each instance was run three times, with averaged results reported.

\begin{figure}[t]
  \centering
  \includegraphics[width=\linewidth]{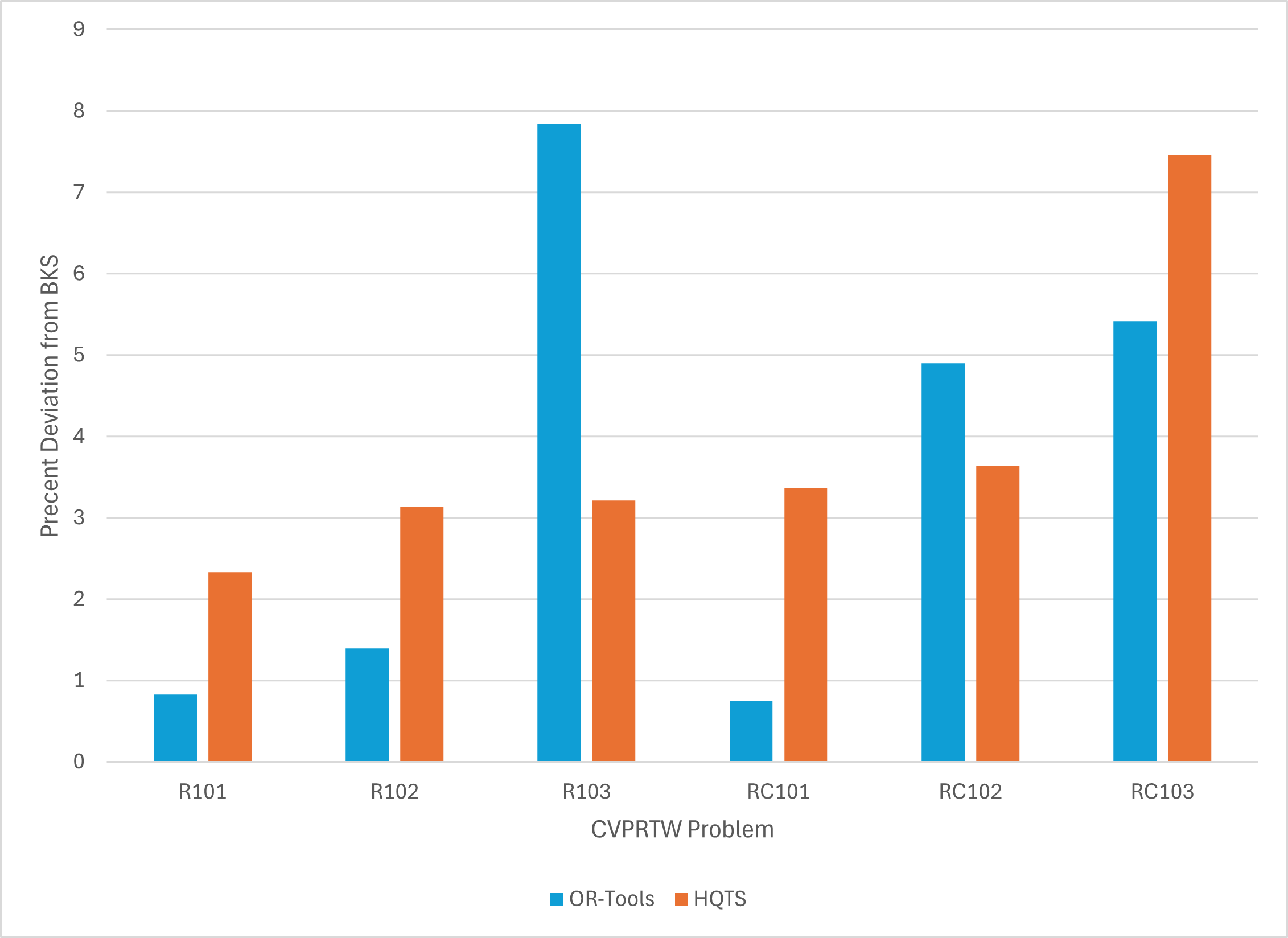}
  \caption{Comparison of Average Percent Deviation from Best Known Solution(s) for HQTS and OR-Tools on the Solomon CVRPTW Subset}
  \label{fig:deviation}
\end{figure}

\begin{figure*}[htbp]
    \centering
    
    \begin{subfigure}[b]{0.32\textwidth}
        \centering
        \includegraphics[width=\textwidth]{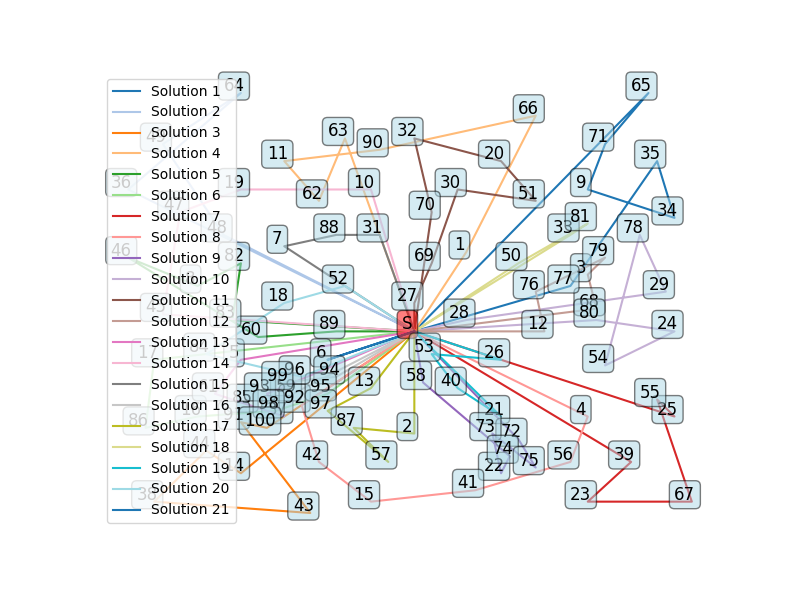}
        \caption{R 101}
        \label{fig:r101}
    \end{subfigure}
    \hfill
    \begin{subfigure}[b]{0.32\textwidth}
        \centering
        \includegraphics[width=\textwidth]{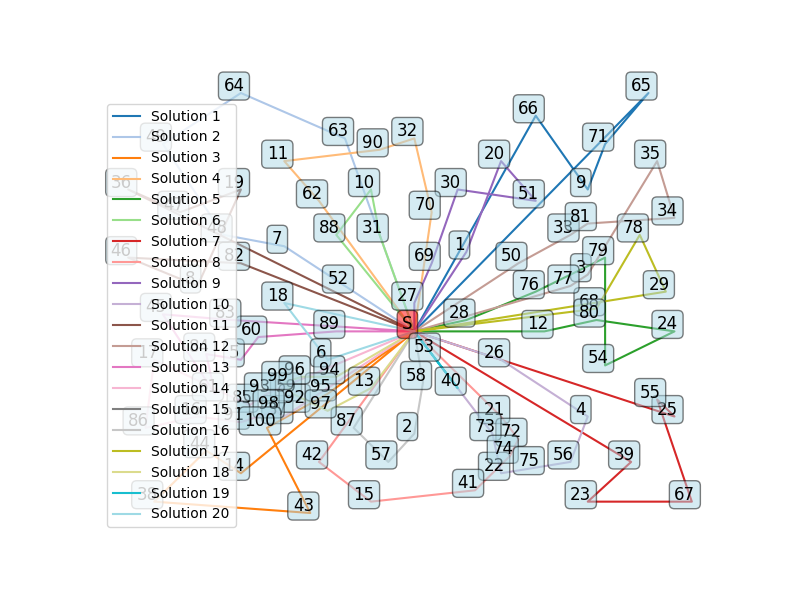}
        \caption{R 102}
        \label{fig:r102}
    \end{subfigure}
    \hfill
    \begin{subfigure}[b]{0.32\textwidth}
        \centering
        \includegraphics[width=\textwidth]{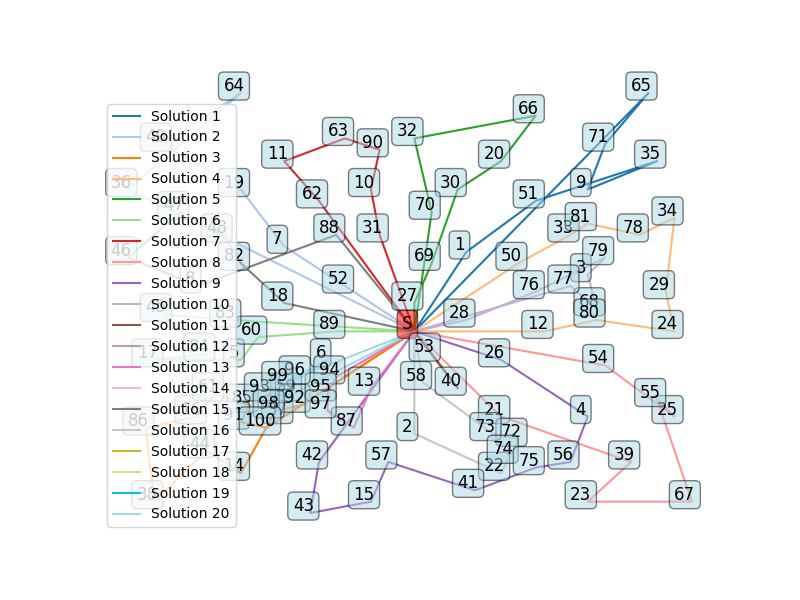}
        \caption{R 103}
        \label{fig:r103}
    \end{subfigure}
    
    \vspace{0.5cm}
    
    \begin{subfigure}[b]{0.32\textwidth}
        \centering
        \includegraphics[width=\textwidth]{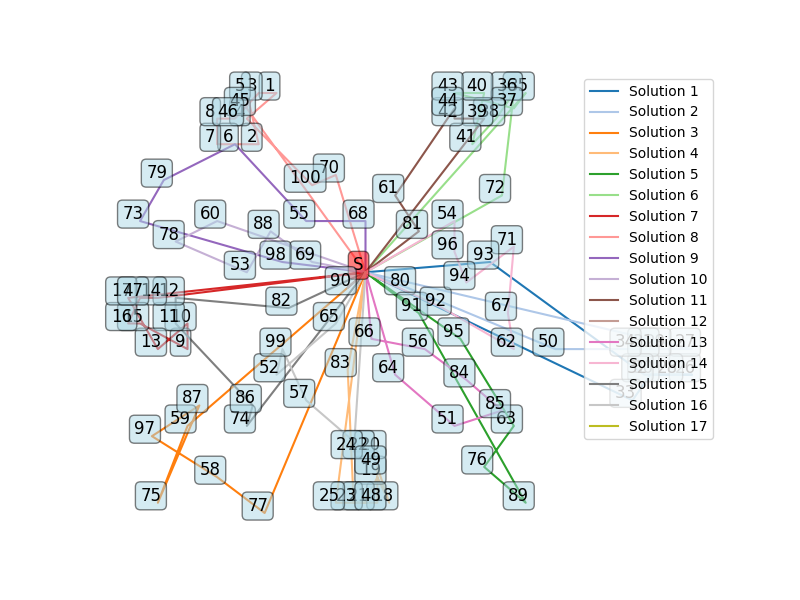}
        \caption{RC 101}
        \label{fig:rc101}
    \end{subfigure}
    \hfill
    \begin{subfigure}[b]{0.32\textwidth}
        \centering
        \includegraphics[width=\textwidth]{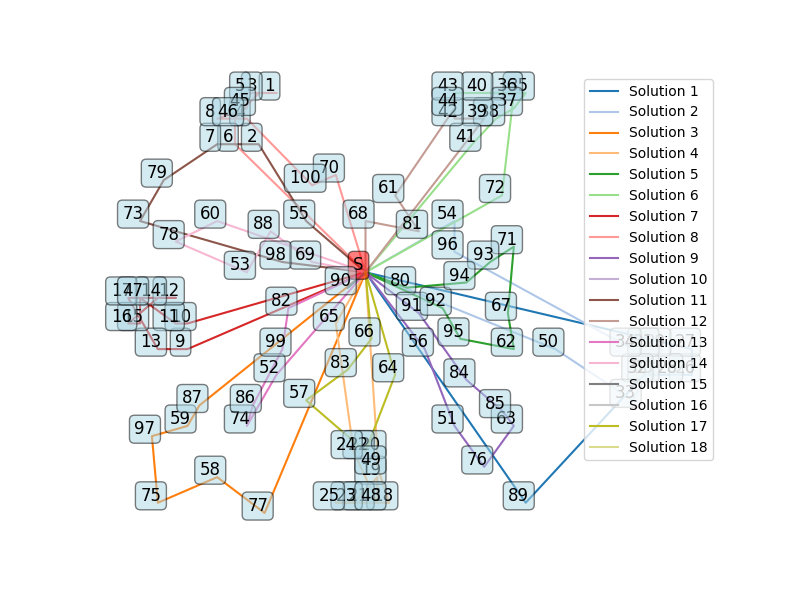}
        \caption{RC 102}
        \label{fig:rc102}
    \end{subfigure}
    \hfill
    \begin{subfigure}[b]{0.32\textwidth}
        \centering
        \includegraphics[width=\textwidth]{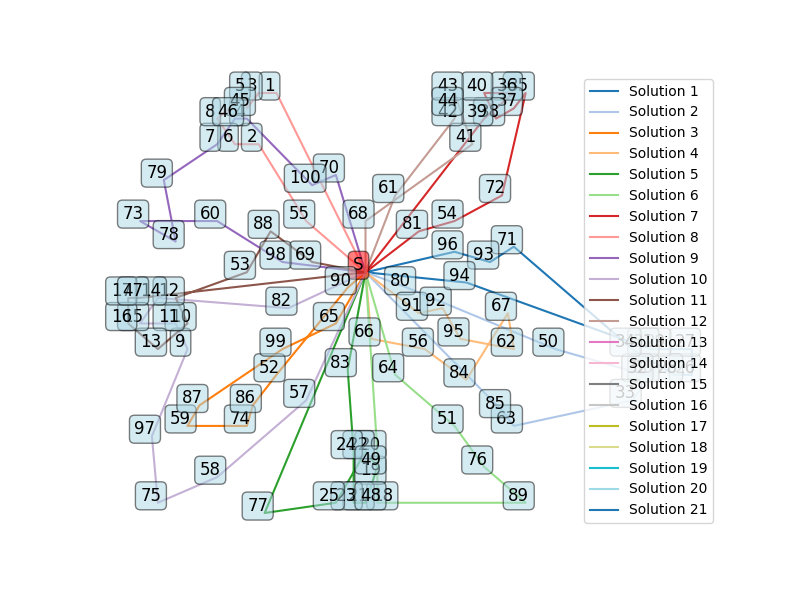}
        \caption{RC 103}
        \label{fig:rc103}
    \end{subfigure}
    
    \caption{HQTS Best Solutions For Solomon Benchmark Dataset}
    \label{fig:overall}
\end{figure*}

To contextualize HQTS’s performance, we benchmarked against Google’s OR-Tools optimization suite~\cite{googleortools,githubortoolsresults}, a robust classical solver, on the same six problems. Figure~\ref{fig:overall} visualizes HQTS’s best solutions, while Table~\ref{hqtsperformance} and Figure~\ref{fig:deviation} detail the comparison. HQTS achieved an average optimality gap of 3.86\% against best-known solutions (BKS), outperforming OR-Tools in two instances (R103: 3.22\% vs. 7.84\%; RC102: 3.64\% vs. 4.89\%) despite OR-Tools’ tighter average gap of 3.52\%. HQTS maintained a consistent trend line as problem complexity increased, so did distance OR-Tools was more sporadic. Still, the performance between the two is similar on average.

\begin{table}[h]
    \caption{Average Performance Comparison of OR-Tools and HQTS on a Solomon CVRPTW Subset}
    \label{hqtsperformance}
    \centering
    \scalebox{0.9}{
        \begin{tabular}{l c c c c c}

            & & \multicolumn{2}{c}{\textbf{OR-Tools}} & \multicolumn{2}{c}{\textbf{HQTS (Ours)}} \\
            \cline{3-4} \cline{5-6}
            {\textbf{Problem}} & {\textbf{BKS}} & & Optimality & & Optimality \\
            & & Distance  & Gap (\%) & Distance & Gap (\%) \\
            \hline
            R101  & 1,637.7 & 1,651.2 & \textbf{0.82} & 1,675.9 & 2.33 \\
            R102  & 1,466.6 & 1,487.0 & \textbf{1.39} & 1,512.6 & 3.13 \\
            R103  & 1,208.7 & 1,303.5 & 7.84 & 1,247.6 & \textbf{3.22} \\
            RC101 & 1,619.8 & 1,632.0 & \textbf{0.75} & 1,674.3 & 3.37 \\
            RC102 & 1,457.4 & 1,528.8 & 4.89 & 1,510.4 & \textbf{3.64} \\
            RC103 & 1,258.0 & 1,326.1 & \textbf{5.41} & 1,351.9 & 7.46 \\
            \hline
        \end{tabular}
    }
\end{table}

\section{Conclusions}\label{conclusion}
Prior work with the D-Wave QA \cite{holliday2024tabu} showcased its robust optimization capabilities for combinatorial problems. However, incorporating constraints such as time windows in the TSPTW often produces infeasible solutions. This inconsistency prompted the development of a swapping heuristic to rectify infeasible outcomes from the CQM solver, integrated into HQTS. Applied to a subset of the Solomon CVRPTW dataset \cite{solomon1987algorithms}, HQTS achieved an average optimality gap of 3.86\%, demonstrating competitive performance despite constraint-induced challenges. The comparison with OR-Tools highlights HQTS’s competitive edge in specific scenarios, leveraging quantum-classical synergy where classical methods falter. However, OR-Tools’ consistency underscores the challenge of quantum feasibility under tight constraints.

Our experiments revealed that while the CQM solver excels at minimizing objective functions, it struggles to enforce time window feasibility, a limitation we did not mitigate by adjusting runtime parameters, an avenue left for future exploration. Benchmarking (HQTS's 3.86\% gap) against Google OR-Tools~\cite{googleortools} revealed a nuanced landscape: while OR-Tools averaged a 3.52\% gap, HQTS surpassed it in R103 and RC102, signaling hybrid QC’s potential to outshine classical solvers in targeted cases. These findings underscore QC's promise for tackling complex routing problems, yet highlight the critical role of hybrid quantum-classical approaches and post-processing in ensuring feasible solutions under stringent constraints. 

Looking ahead, we aim to extend our evaluation across a broader, more diverse CVRPTW dataset, varying customer counts and problem scales. Furthermore, we intend to refine hybrid QC strategies, building on frameworks like HQTS, to address real-world optimization challenges such as the CVRPTW.

{
    \small
    \bibliographystyle{IEEEtran}
    \bibliography{cvrptw}
}

\end{document}